\documentclass[acmsmall,screen]{acmart}
\AtBeginDocument{%
  }

\setcopyright{acmlicensed}
\copyrightyear{2018}
\acmYear{2018}
\acmDOI{XXXXXXX.XXXXXXX}

\acmConference[Conference acronym 'XX]{Make sure to enter the correct
  conference title from your rights confirmation emai}{June 03--05,
  2018}{Woodstock, NY}
\acmISBN{978-1-4503-XXXX-X/18/06}

\usepackage{enumitem}
\usepackage[most]{tcolorbox}  
\usepackage{xspace}

\newcommand{\xiuhu}{CCF Beautiful Lake Seminars\xspace}

\begin{document}

\title{The Current Challenges of Software Engineering in the Era of Large Language Models}

\author{Cuiyun Gao}
\email{gaocuiyun@hit.edu.cn}
\affiliation{%
  \institution{Harbin Institute of Technolgy}  
  \country{China}
}

\author{Xing Hu}
\email{xinghu@zju.edu.cn}
\authornote{Corresponding Authors: Xing Hu and Zhi Jin.}
\affiliation{%
  \institution{Zhejiang University}
  \country{China}
  }

\author{Shan Gao}
\email{gaoshan17@huawei.com }
\affiliation{%
  \institution{Huawei Technologies}
  \country{China}
  }

\author{Xin Xia}
\email{xin.xia@acm.org}
\affiliation{%
 \institution{Huawei Technologies}
 \country{China}
}

\author{Zhi Jin}
\email{zhijin@pku.edu.cn}
\authornotemark[1]
\affiliation{%
  \institution{Peking University}
  \country{China}
}

\renewcommand{\shortauthors}{Gao et al.}

\begin{abstract}
With the advent of large language models (LLMs) in the artificial intelligence (AI) area, the field of software engineering (SE) has also witnessed a paradigm shift. These models, by leveraging the power of deep learning and massive amounts of data, have demonstrated an unprecedented capacity to understand, generate, and operate programming languages. They can assist developers in completing a broad spectrum of software development activities, encompassing software design, automated programming, and maintenance, which potentially reduces huge human efforts. Integrating LLMs within the SE landscape (LLM4SE) has become a burgeoning trend, necessitating exploring this emergent landscape's challenges and opportunities. 

The paper aims at revisiting the software development life cycle (SDLC) under LLMs, and highlighting challenges and opportunities of the new paradigm. The paper first summarizes
the overall process of LLM4SE, and then elaborates on the current challenges based on a through discussion. The discussion was held among more than 20 participants from academia and industry, specializing in fields such as software engineering and artificial intelligence. Specifically, we achieve
26 key challenges from seven aspects, including software requirement \& design, coding assistance, testing code generation, code review, code maintenance, software vulnerability management, and data, training, and evaluation. We hope the achieved challenges would benefit future research in the LLM4SE field.
\end{abstract}




\maketitle

\section{Introduction}
In the 1960s, software engineering (SE) was formally established as a discipline in response to software's increasing scale and complexity~\cite{DBLP:conf/icse/Royce87}, based on the symposium on SE organized by the North Atlantic Treaty Organization (NATO). In the nascent stage of SE, during which artificial intelligence (AI) techniques were not widely adopted, the field was more focused on structured programming~\cite{DBLP:books/daglib/0070529}, modular design~\cite{stone1997towards}, and data structure~\cite{hopcroft1983data}, which greatly promoted the widespread application of programming languages and compiler systems. Along with the explosive development of object-oriented programming (OOP) and cloud computing, and continuing maturity of the AI techniques, the field of intelligent software engineering (also called \textit{AI4SE})~\cite{DBLP:conf/icse/Harman12} has gradually emerged. In the field, AI techniques have been introduced into various stages of the software development lifecycle (SDLC), including programming, testing, and maintenance, aiming at boosting the efficiency and quality of software development. Nowadays, AI4SE is still playing an increasingly essential role in a growing number of SE domains, including but not limited to intelligent requirements analysis and design~\cite{DBLP:conf/icse/ParraVA18,DBLP:conf/icse/BusariL17}, automated code generation and repair~\cite{DBLP:conf/icse/LiWLWCWG23,DBLP:conf/icse/JiangL021,DBLP:conf/icse/MuCSWW23}, as well as intelligent project management~\cite{DBLP:journals/spe/AkbarKIM24, DBLP:journals/icae/LinDGHK15}. 
According to Gartner's report for 2023~\cite{gartner}, artificial intelligence-augmented software engineering is listed as one of the top strategic technology trends in SE, indicating the general trend of enhancing SE with AI techniques.

Recently, the rapid advancement of the Artificial Intelligence Generated Content (AIGC) technology~\cite{DBLP:journals/corr/abs-2303-04226}, especially large language models (LLMs)~\cite{34,DBLP:conf/acl/ZanCZLWGWL23}, is ushering in new challenges and opportunities for the SE field, and displays potential for transforming the software development paradigm. LLM-based software engineering (also called \textit{LLM4SE}) has attracted widespread attention from both academia and industry. The practical application of tools such as GitLab Duo~\cite{GitLab} and GitHub Copilot X~\cite{copilot} have demonstrated the broad prospects of LLM4SE, inspiring new software development mode. {Specifically, by}
breaking through the traditional multi-stage software development process, LLM4SE would make an end-to-end software development possible, i.e., generating and testing code directly based on given requirements.

The prospects of LLM4SE have acted as the ``catfish effect'', promoting the exploration of new software development modes among academia and industry. 
However, due to limitations such as the essential semantic gap between code and natural languages, constantly updated data and models, and instability of generated results, existing LLM4SE techniques are still faced with serious challenges regarding the quality, efficiency, reliability, and trustworthiness of developed software. 
The challenges have also sparked a new wave of enthusiasm for SE technologies. 
To better understand the issues in the current SE stage, 24 academia researchers and industry practitioners{, specializing in different fields such as software engineering and artificial intelligence,} were grouped to discuss the opportunities and challenges in the era of LLM4SE from 19 Jan 2024 to 21 Jan 2024 at the ``9th \xiuhu''~\cite{ccfxiuhulink}. 
Based on the rigorous discussion, the paper summarizes 26 key challenges from seven aspects, including software requirement \& design, coding assistance, testing code generation, code review, {software maintenance}, software vulnerability management, and data, training, and engineering. This paper aims to provide insights for researchers and practitioners to effectively harness LLMs' strengths while mitigating their weaknesses, thereby applying them to support continuous and robust software development.

The major contributions of this paper are summarized as follows:

\begin{enumerate}
\item We perform a qualitative study to investigate {and summarize} the current challenges of LLM4SE.

\item We {achieve findings from various aspects in LLM4SE, and} provide practical implications for researchers and practitioners to inspire future avenues of research.
\end{enumerate}

{The remainder of this paper is structured as follows: Section~\ref{sec:background} provides the background of our work, including the overall process of LLM4SE and common techniques in each step. Section~\ref{sec:method} describes our methodology for investigating the challenges in LLM4SE. Section~\ref{sec:challenges} and Section~\ref{sec:literature} present the summarized challenges and related work, respectively. Section~\ref{sec:threats} discusses the threats to the validity of our work, and Section~\ref{sec:con} concludes the paper.}

\section{Background}\label{sec:background}

In this section, we first illustrate the overall process of LLM4SE, and then elaborate on the details in each step including the data construction, fine-tuning techniuqes, SE-specific LLMs, prompt tuning techniques, and downstream tasks.

\subsection{Overall Process of  LLM4SE}

\begin{figure*}[htbp]
	\centering
\includegraphics[width=1\textwidth]{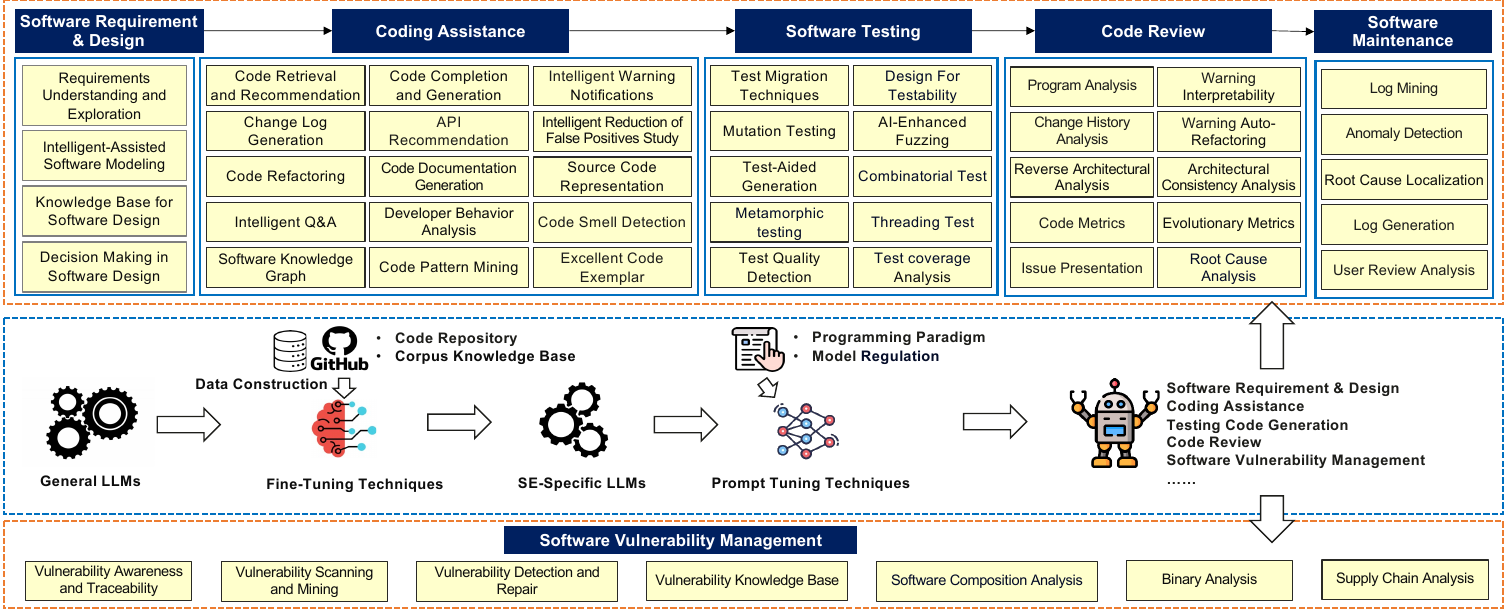}
	\caption{Overall process of LLM4SE.}
	\label{workflow}
\end{figure*}

Figure~\ref{workflow} presents the overall process of the integration of LLM into software engineering applications, including data construction, fine-tuning, prompting techniques, and Software Engineering (SE)-specific LLMs. With the support of SE-specific LLMs, multiple SE tasks (e.g., design, coding, testing, reviewing, maintenance, and vulnerability management) can be enhanced substantially. 

\subsection{Data Construction}
The remarkable performance of the previously mentioned large code models can be attributed to the extensive dataset and the meticulously structured data organization employed during the model's pre-training phase.

\subsubsection{Data Collection}
Data collection is the basic step for constructing a high-quality corpus of source code. A common solution is to collect publicly available source code data from GitHub or public archive~\cite{lozhkov2024starcoder}. For instance, Guo et al.~\cite{23} crawl public repositories created before February 2023 on GitHub with 87 programming languages. The StarCoder2 team collects the source code from the Software Heritage (SH) archive~\cite{abramatic2018building}. Then the authors start by extracting the most recently crawled versions of all GitHub repositories and filtering them to retain only the main branch. 

\subsubsection{Data Filtering}
Source code data from public repositories may not always be clean or useful, and malicious data may harm model performance and leak confidential user data. Therefore, delicate data filtering is also crucial for powerful large code models. According to Lozhkov et al.~\cite{lozhkov2024starcoder}, many filtering rules are applied for StarCoder2, including 1) \textbf{Long line filters} to eliminate too long code lines, 2) \textbf{Auto-generated filter} to remove auto-generated files such as compilation files, 3) \textbf{Alpha filter} to remove files with less than 25\% of alphabetic characters, and 4) \textbf{Encoded data filter} to remove content about encoding data.

\subsubsection{Data Organization}
According to the work~\cite{23}, there exists an inter-relationship among the files of source code. Thus, the authors propose a Topological Sort for Dependency Analysis algorithm for organizing code files. Based on the dependencies, models are trained with the files in a certain order to learn such dependencies. For StarCoder2 \cite{lozhkov2024starcoder}, authors utilize different special tokens such as $<commit>$ to represent different data types, such as commit messages.

\subsection{Fine-Tuning Techniques}
Through pre-training, LLMs learn a huge amount of knowledge about software engineering and source code. However, LLMs are not aligned with human preference, resulting in their limited capabilities to serve as intelligent assistants. 
{Therefore, to better adapt LLMs for software development practice, researchers primarily focus on four tuning techniques: adapter-based tuning, low-rank adaptation, prefix tuning, and prompt tuning.}
{\subsubsection{Adapter-based tuning.} Adapter-based tuning introduces adapter modules with a bottleneck structure between the layers of pretrained models. In the fine-tuning stage, the parameters of the original pretrained models are fixed, and only adapter modules are adjusted.
In particular, Wang et al.~\cite{DBLP:conf/icse/WangCLLPDL23} introduce adapter tuning to improve code search and summarization by adding a small number of trainable parameters to pre-trained models. Their approach reduces catastrophic forgetting in LLM and achieves superior performance in cross-lingual and low-resource scenarios.}
{\subsubsection{Low-rank adaptation.} Low-rank adaptation injects trainable rank decomposition matrices into pretrained models to substitute the original weights, which can amplify some hidden features encoded during pre-training. In Liu et al.’s study~\cite{DBLP:conf/wcre/LiuKYLZL24}, they explore how Low-Rank Adaptation (LoRA) modifies pre-trained models for code-change tasks by injecting low-rank matrices that substitute original weights. Through this method, they effectively captures dynamic code changes while reducing the need for extensive retraining.}
{\subsubsection{Prefix tuning.} Prefix tuning introduces a set of prefix parameters to the beginning of each layer in the LLM. These parameters are serve as adjustments that precede the original input during processing. LLaMA-Adapter~\cite{DBLP:conf/iclr/ZhangHLZL00024} incorporates a set of trainable adaptation embeddings and attaches them to the word embeddings in the upper layers of the LLMs. A zero-initialized attention scheme with zero gating is also introduced. It dynamically incorporates new guiding signals into LLaMA while retaining its pretrained knowledge.}
{\subsubsection{Prompt tuning.} Prompt tuning modifies an LLM's input by incorporating a natural language prompt designed to predict masked tokens. This adjustment maintains the input format consistent with that used during the model's pre-training, facilitating more effective task-specific adjustments. 
For instance, Wang et al.~\cite{DBLP:conf/sigsoft/WangYGP0L22} apply prompt tuning to pre-trained models like CodeBERT and CodeT5 and showing its effectiveness over fine-tuning in various code intelligence tasks at low-resource scenario. The inserted natural language prompt can involve task-specific knowledge to facilitate the adaption to downstream tasks.}

\subsection{SE-Specific LLMs}
At the end of 2021, OpenAI released CodeX~\cite{chen2021evaluating} as the first large language model designed explicitly for code generation. With a similar architecture to GPT-3~\cite{brown2020language}, CodeX was pre-trained using 159 GB of public code data from GitHub. Based on CodeX, the Copilot plugin~\cite{copilot} has become one of the most popular tools for code generation. Additionally, the HumanEval dataset proposed in the work ~\cite{chen2021evaluating} has become one of the standard evaluation datasets for subsequent code generation tasks. In 2022, DeepMind proposed AlphaCode~\cite{3}, a model that employed an encoder-decoder architecture and was pre-trained on 715.1GB of code data across 12 programming languages. It surpassed 54.3\% of human programmers in a CodeForce competition with 5000 participants. 

In recent years, a growing number of open-source SE-Specific LLMs have been proposed. In 2020, Microsoft proposed CodeGPT~\cite{svyatkovskiy2020intellicode}, based on the architecture of GPT-2, which was used to support the embedded code completion tool of Visual Studio. Wang et al. proposed CodeT5~\cite{29} that follows an encoder-decoder architecture to support various code intelligence tasks, including code translation and code summarization. Salesforce proposed CodeGEN~\cite{13}, an open-source model with 1.61 billion parameters, pre-trained on over 800GB of data. Unlike other models that generate code in a single step, CodeGEN facilitates conversational program generation. The process of code generation with the help of the model can be seen as a multi-round dialogue between the user and the model, where the user provides requirements for the code generation model in multiple instances and receives several outputs from the model.

In addition, Meta AI proposed CodeLlama, a family of code LLMs based on LLama 2~\cite{codellama} with the capabilities of code generation, blank infilling, and long-context processing. The BigCode project introduced StarCoder~\cite{22}, a large language model trained on the mixture of source code and natural language texts. Its training data incorporate more than 80 different programming languages, as well as text extracted from GitHub issues and commits and from notebooks. The total account of training tokens exceeds 1T. Based on CodeLlama, Guo et al. proposed DeepSeek-Coder~\cite{23}, which is a series of code LLMs that have an identical architecture to CodeLlama. DeepSeek-Coder is trained from 2T tokens from scratch, which comprises 87\% code and 13\% natural language in both English and Chinese. DeepSeek-Coder has achieved state-of-the-art performance in various code intelligence tasks, e.g., code generation and code completion. Lozhkov et al. proposed StarCoder2~\cite{lozhkov2024starcoder} and a larger-scale source code dataset named The Stack V2. Based on purely open-sourced data, StarCoder2 achieves state-of-the-art effectiveness in math and code reasoning benchmarks, as well as several low-resource languages. The aiXcoder team has launched aiXcoder-7B~\cite{DBLP:journals/corr/abs-2410-13187}, a lightweight and effective LLM for code completion. More LLMs for SE are illustrated in Table~\ref{tab:codellm}.

\subsection{Prompting Techniques}
Prompting techniques are the key factors in adapting LLMs into SE practice. In this section, we categorize the prompting techniques into two types including training-time and test-time techniques.

\textbf{Training-time techniques}. Wang et al. \cite{DBLP:conf/sigsoft/WangYGP0L22, wang2023prompt} observe that prompt tuning that is equipped with the templates describing task formulation and specification can further boost the performance in downstream tasks compared to fine-tuning. Luo et al. \cite{46} utilize the prompting techniques Evol-Instruct \cite{xu2023wizardlm} to construct a more complex dataset and adopt a delicately designed template to describe the task in the training process. 

\textbf{Test-time techniques}. A direct way to use LLMs in SE tasks is \textit{zero-shot prompting}, which describes the task and obtains answers from LLMs. To make LLMs further understand the task, \textit{few-shot prompting} is proposed \cite{brown2020language} (is also called in-context prompting). Before asking the question, multiple input-output pairs are selected as examples to present to LLMs. With the demonstrated examples, LLMs can further understand the task specification and output format, producing better performance \cite{DBLP:conf/kbse/GaoWGWZL23}. In addition, to further specify the task and improve the reasoning capabilities of LLMs, Wei et al. propose \textit{chain-of-thought} \cite{wei2022chain} (COT) that prompts LLMs the detailed reasoning process instead of directly showing the answer, substantially improving the performance of LLMs. Equipped with COT, LLMs have also been proven to present a better performance in SE tasks such as code generation \cite{23}.

\subsection{Downstream Tasks}

{The LLM4SE workflow can be applied to various SE tasks, including software requirement \& design, coding assistance, testing code generation, code review, software maintenance, and software vulnerability management. Specifically, for the software requirement \& design task, LLMs can serve as a knowledge base for software design, or conduct code-centric software design, etc. For the coding assitance, LLMs are beneficial for a series of tasks, including code retrieval and recommendation, code pattern mining, code completion and generation, and intelligent Q\&A. For testing code generation, LLMs could support threading test, test coverage analysis, and AI-enhanced fuzzing, etc. For the software maintenance, LLMs are able to support many tasks such as log mining, anomaly detection, and user review analysis. Finally, for the software vulnerability management, LLMs can facilitate vulnerability awareness and traceability, binary analysis, and supply chain analysis, etc.}

\begin{table*}
\centering
\caption{\label{table_1}LLMs for software engineering. The symbol ``-'' indicates the corresponding value is not publicly accessible.}\label{tab:codellm} \vspace{-0.3cm}\small
\resizebox{\linewidth}{!}{
\begin{tabular}{cccccc} 
\hline
Model & Developer & Release Date & Parameter Quantity (B) & Pre-trained Data Size& Architecture \\ \hline

CodeX \cite{chen2021evaluating} & OpenAI & 2021.07&2.5/12&100B tokens&decoder\\
AlphaCode \cite{3} &DeepMind&2022.02&41&967B tokens&encoder-decoder\\
CodeGen \cite{13}&Salesforce&2022.03&2.7/6.1/16&577B tokens&decoder\\
AiXcoder L&AiXcoder&2022.04&-&-&decoder\\
AiXcoder XL \cite{4} &AiXcoder&2022.06&-&-&decoder\\
PanGu-Coder \cite{8} &HUAWEI&2022.07&2.6&147GB&decoder\\
CodeGeeX \cite{9} &Zhipu.AI&2022.09&13&850B tokens&decoder\\
BLOOM \cite{10} & BigSicence&2022.09&176&1.61 TB&decoder\\
PaLM-Coder \cite{6} & Google&2022.10&8/62/540&780B tokens&decoder\\
SantaCoder \cite{20} &Huggingface&2023.01&1.1&118B tokens&decoder\\
InCoder \cite{17} &Facebook&2023.04&1.3/6.7&216 GB&decoder\\
StarCoder \cite{22} &Huggingface&2023.05&15.5&1T tokens&decoder\\
CodeGen2 \cite{19} &Salesforce&2023.05&3.7/7/16&400B tokens&decoder\\
CodeT5+ \cite{wang2023codet5+} &Salesforce&2023.05&2/6/16&51.1B tokens&encoder-decoder\\
WizardCoder \cite{46} &Microsoft&2023.06&15&-&decoder\\
Phi-1 \cite{phi1} &Microsoft&2023.06&13&7B tokens&decoder\\
PanGu-Coder2 \cite{pangucoder2} &HUAWEI&2023.07&15&-&decoder\\
CodeLLaMa \cite{codellama} &Meta&2023.08&7/13/34&116K tokens&decoder\\
{CodeQwen} \cite{codeqwen}& Alibaba & 2023.09 & 7/14 & 90B tokens & decoder \\
Phi-1.5 \cite{phi15} &Microsoft&2023.09&1.3&27B tokens&decoder\\
DeepSeek-Coder \cite{23} &DeepSeek AI&2023.11&1/5.7/6.7/33&2T tokens&decoder\\
Phi-2 & Microsoft &2023.12&2.7&1.4T tokens&decoder\\
AlphaCode2 & DeepMind &2023.12&-&-&decoder\\
Magicoder \cite{wei2023magicoder} & UIUC\&THU&2023.12&7&-&decoder\\
WaveCoder \cite{yu2023wavecoder} & Microsoft&2024.01&15&-&decoder\\
StarCoder2 \cite{lozhkov2024starcoder} & Huggingface &2024.01& 3/7/15 & 622B+/658B+/913B+ tokens&decoder\\
{CodeQwen1.5} & Alibaba & 2024.04 & 7 & 3T tokens &  decoder \\
{SEMCODER} \cite{semcoder} & Columbia University & 2024.06 & 6.7 & - & decoder \\
{DeepSeek-Code-V2} \cite{deepseekcoder2} & DeepSeek AI & 2024.06 & 16/236 & 10.2T tokens & MoE \\
{Qwen2.5-Coder} \cite{codeqwen2.5} & Alibaba & 2024.09 & 1.5/7/32 & 5.5T tokens  & decoder  \\
\hline
\end{tabular}
}
\end{table*}
\section{Methodology}\label{sec:method}

To investigate the challenges in LLM4SE, 24 participants with experience in software development had face-to-face meetings in the ``9th \xiuhu'' during 19 Jan 2014 - 21 Jan 2024. {The participants include 17 academic researchers and 7 industry practitioners, specilizing in different fields such as software engineering and machine learning}. Six thematic sessions were performed, namely ``Software Engineering under Large Models'', ``Large Models for Code Intelligence'', ``Large Models for Software Quality Assurance'', ``Data, Evaluation, and Validation for Large Code Models'', ``Large Models for Open-source Engineering'', and ``Future Trends and Challenges of LLM4SE'', each followed by a panel or collective discussion{, with detailed topics illustrated in the public repository\footnote{\url{https://docs.google.com/spreadsheets/d/1ewVENYYq1UyKBQDaAgC8-45WMI8fNCREJdlGHgRdGbg/edit?usp=sharing}}}. Each thematic seminar together with the discussion lasted for around four hours for ensuring the corresponding topics were thoroughly discussed. We performed the following coding procedures to summarize the discussed challenges.

\textbf{(1) Transcribing and Coding.} {Following the procedure in the work \cite{DBLP:conf/icse/HuX0WCZ22,DBLP:conf/sigsoft/WangHGJ0HLD23}, the first author transcribes the discussion of each seminar and uses NVivo \cite{edwards2014qualitative} to conduct open coding and generate opinion cards. Then another author verifies the summarized opinion cards and provides suggestions for improvement.}


\textbf{(2) Open Card Sorting.} The two authors then separately grouped the codes into potential challenges for thematic similarity according to LaToza et al.'s study~\cite{DBLP:conf/icse/LaTozaVD06}. {The two
authors discuss their disagreements to reach a common decision. To reduce the bias of the two authors in sorting descriptions, another two authors have also reviewed and confirmed the final set of challenges.}
Finally, we derived 26 key challenges corresponding to seven aspects, with details illustrated in Section~\ref{sec:challenges}.

\section{Current Challenges  in LLM4SE}\label{sec:challenges}
In this section, we present the challenges of LLM4SE. We introduce them from six aspects, including the challenges in the software engineering process, e.g., software requirement and design, coding assistance, testing, code analysis and review, software vulnerability management, as well as the challenges relevant to the construction of SE-specific LLMs, e.g., challenges in data, training, and evaluation.

\subsection{Challenges in   Requirement \& Design}
Requirement~\cite{DBLP:conf/icse/NuseibehE00} and design~\cite{DBLP:journals/tse/YauT86} constitute critical phases in the software development process, and serve as initial stages
of the process. The primary challenges stem from two main aspects, including the inherent complexity of real-world problems and communication barriers between domain experts and software developers. {These two aspects
lead to increased uncertainty in defining requirements during the early stages of development, and thereby unexpected design. The powerful capabilities of LLMs in natural language processing can effectively mitigate expression uncertainties, thereby beneficial for addressing the challenges encountered during the requirements and design stages.}
However, how to ensure that LLMs can effectively capture the intricate requirements and design in the complex software still remains challenging.

\noindent\textbf{(1) Requirement/design prompts.} {LLMs heavily rely on comprehensive prompts and available contextual information to generate effective outputs. Even slightly different prompts can lead to significantly different results~\cite{DBLP:conf/sigsoft/WangYGP0L22}. Besides, it is hard to render the prompts anticipate the costs and risk of the generated requirements/design~\cite{DBLP:journals/corr/abs-2303-07839}.} Therefore, when employing LLMs as requirement and design agents, it is essential to conduct thorough empirical evaluations on the prompts.


\noindent\textbf{(2) Structured descriptions of the communication between domain experts and software developers.} LLMs present inherent limitations in context length and inductive bias (influenced by training data)~\cite{levine2021inductive}. For instance, the context length of ChatGPT (e.g., gpt-3.5-turbo) is restricted to 16,385 tokens, which is usually sufficient but still hard to process lengthy requirement documents or maintain the task context over extended conversations.

\noindent\textbf{(3) Lack of domain expert knowledge.} Requirements engineers are generally required to comprehend the application domain for accurate and complete requirement extraction. However, LLMs may have limited training on specific domain knowledge, necessitating the integration of domain knowledge through experts, or fine-tuned LLMs tailored for the domain. {Without clear software requirements, it would be challenging to produce accurate software design.}

\noindent\textbf{(4) Evolvability of software requirements.} The evolution of requirements is the origin of software evolution. Prior studies ~\cite{DBLP:conf/sigsoft/BlincoeVD13,DBLP:conf/kbse/YangWRN10} more focus on fixed requirements, ignoring the evolvability of software requirements. How to incorporate LLMs into the adaptive perception and generation of evolving requirements has not received attention.

\noindent{\textbf{(5) Comprehensive evaluation of generated requirements and design.} As defined in ISO 29148~\cite{iso2015iec}, the quality of software requirements should be measured in various characteristics, including completeness, correctness, verifiability, unambiguousness, singularity, and feasibility, etc~\cite{DBLP:conf/re/LubosFTGMEL24}. The generated software design is expected to include concrete design specifications, domain-specific languages, and explore alternative architectures~\cite{DBLP:journals/corr/abs-2303-07839}. How to automate the evaluation of LLM-based software requirements and design process has received scant attention.}

\noindent\textbf{(6) Inconsistency between software modeling and natural language descriptions.} Prior studies~\cite{DBLP:journals/sosym/CamaraTBV23,DBLP:journals/corr/abs-2410-17370, DBLP:conf/models/ChenYCLMV23,DBLP:conf/models/ArulmohanM023} have demonstrated the software modeling capabilities of LLMs, such as generating Unified Modeling Languages (UMLs) based on user stories, classifying model repositories, etc. Although LLMs can generate syntactically correct UML code, they are struggling with semantic quality and tend to align poorly with natural languages, especially when the diagram scale is large.

\begin{tcolorbox}[breakable,width=\linewidth-2pt,boxrule=0pt,top=3pt, bottom=3pt, left=4pt,right=4pt, colback=gray!15,colframe=gray!15]
\textbf{Summary:} The area of LLM for requirement and design still faces several challenges, including effective requirement/design prompts, structured descriptions of the communication between domain experts and software developers, lack of domain knowledge, dynamics of software requirements, comprehensive evaluation, and inconsistency in software modeling. 
\end{tcolorbox}

\subsection{Challenges in Coding Assistance}

During the software development cycle, the advent of code generation technologies powered by LLMs has greatly improved development efficiency and code quality, allowing developers to address business challenges instead of engaging in monotonous coding tasks. Despite the opportunities these advanced models introduce to code generation, they also present notable challenges. 

\vspace{0.1cm}\noindent \textbf{(1) Inaccurate generated code caused by hallucination in LLMs.} Even the most advanced LLMs may generate results that contradict the facts {\cite{chelli2024hallucination}}, compromising their reliability in code generation applications. This problem becomes especially critical in high-risk settings where strict compliance with specific application requirements is essential. The potential for inaccuracies in these environments underscores the substantial risks of deploying such models in production scenarios.

\vspace{0.1cm}\noindent \textbf{(2) Introduction of vulnerabilities in generated code.} The pre-training datasets for LLMs are primarily from open-source code repositories. Contributors to these repositories come from different backgrounds and possess varying levels of expertise, leading to numerous security vulnerabilities in the code. When LLMs learn from such datasets, they may unintentionally replicate these security issues in their outputs {\cite{wang2024your}}. Deploying models trained on this data in real-world development projects can inadvertently introduce these vulnerabilities, posing risks with unforeseen consequences. 

\vspace{0.1cm}\noindent \textbf{(3) Limited generation performance for new programming languages.} The advent of emerging programming languages such as  Rust presents a challenge due to the scarcity of available training data. Despite an urgent need for programming assistants to lower the barrier to entry and promote these new languages, the contradiction between the scarcity of training data and the pressing need for promotion poses a considerable challenge in creating effective LLMs-based code generation assistants.

\vspace{0.1cm}\noindent \textbf{(4) Insufficient evaluation methods for machine-generated code.} Currently, the evaluation of code generation models primarily utilizes datasets like HumanEval {\cite{chen2021evaluating}} and MBPP {\cite{austin2021program}}, characterized by their simple and direct test cases. However, real development scenarios demand that models handle complex real-world code involving interactions across multiple methods and files to complete tasks {\cite{ding2024crosscodeeval}}. To effectively evaluate the capabilities of LLMs in code generation, it is crucial to incorporate this complexity of real-world development environments.

\vspace{0.1cm}\noindent \textbf{(5) Efficient integration of the generated code into projects.} Currently, most of the code generated by an LLM can solve common coding problems, e.g., algorithm problems. However, when given a coding task (e.g., adding a new feature or fixing a bug) in a real project, developers need to first identify the locations where they need to write/modify the code, and then complete the project-specific code {\cite{zhang2023repocoder}}. Note that the newly added code should follow the coding style of the current project and call relevant APIs within the project. It is challenging for LLMs to well capture the project-level knowledge, remember the relevant APIs, identify the coding positions, and provide maintainable code with same coding styles. 

\vspace{-0.3cm}\begin{tcolorbox}[breakable,width=\linewidth-2pt,boxrule=0pt,top=3pt, bottom=3pt, left=4pt,right=4pt, colback=gray!15,colframe=gray!15]
\textbf{Summary:} For code generation, it needs to overcome numerous challenges posed by LLMs, such as issues with accuracy of generation, security vulnerabilities, insufficient support for new programming languages, inadequacies in evaluation methods, and efficient integration of the generated code into projects, to ensure the effective and safe application of LLMs.
\end{tcolorbox} \vspace{-0.3cm}

\subsection{Challenges in Testing Code Generation}

The remarkable performance capability of LLMs offers new opportunities for research in software testing, especially unit test generation. 
The industry currently utilizes LLMs for unit test generation mainly involving two steps: first combining the method under test with its code context into a prompt for querying LLMs, and then refining the LLMs' output through text post-processing to ensure the successful compilation and execution.

The challenges for high-quality test cases are twofold, encompassing syntactics and semantics aspects. Specifically, at the syntactics level, the test code must be compiled and executed correctly. At the semantic level, the test code should comprehensively cover the tested code and support defect detection throughout the process in SDLC for maintaining code quality. Although LLMs can generate test code currently~\cite{DBLP:conf/icse/NashidSM23,DBLP:journals/corr/abs-2305-04207}, providing high-quality test cases that satisfy both syntactical correctness and comprehensive semantic coverage remains a considerable challenge.

\vspace{0.1cm}\noindent \textbf{(1) The syntactics-related challenge.} 
In unit testing, the generated test code often contains code elements that cannot be successfully parsed by compilers {\cite{siddiq2024using}} (e.g., undeclared variable types or non-existent identifiers) and non-code texts (e.g., explanatory natural language text). This limits their practical application in software development, as developers cannot directly utilize these outputs. Therefore, the challenge is to guide or constrain LLMs to reduce texts that cause compilation errors and to enhance the syntax compilation success rate of generated test cases. Two solutions to address this challenge include model fine-tuning (Supervised Fine-Tuning, SFT) {\cite{shin2024domain}}, which manually collects and annotates test code with specific formats to fine-tune the model for satisfying syntax requirements, and prompt engineering, which involves instructions in the prompts to guide the model not to output irrelevant code/texts. However, developing these techniques to keep pace with the evolving capabilities of LLMs remains a significant challenge.

\vspace{0.1cm}\noindent \textbf{(2) The semantics-related challenge.} First, when ensuring that the test code can be compiled successfully, we then need to inspect whether the test cases can cover all executable paths of the code under test. To achieve comprehensive path coverage, test cases are expected to be diverse. This involves specifying preconditions such as inputs, external dependencies, and environmental states for the method being tested. For analyzing these preconditions, various testing techniques can be considered, such as boundary value testing, equivalence partitioning testing, and random testing. Accurately extracting or analyzing these preconditions poses a challenge, which could be addressed by using program analysis techniques to obtain a set of constraints for the preconditions. Another challenge is appropriately assembling these preconditions into prompts {\cite{xie2023chatunitest}} to guide the model in outputting test cases with high coverage. Exploring various prompt engineering techniques could provide solutions for incorporating preconditions effectively.

Second, for large-scale projects with many dependencies, developers frequently mock the objects to simulate the behaviour of the real objects. Thus, when applying SE-specific LLMs to generate test cases, it is challenging for the LLMs to identify the mocking objects automatically, and prefill the data into the mocked objects to simulate the practical behavior. 

Third, high-quality code is characterized not only by correct implementation logic but also by returning correct results, appropriately modifying relevant variables, and capturing and handling exceptions, all in alignment with the intended code design. Unit test cases use assertion statements to verify the expected correctness of the code under test. However, providing LLMs with necessary information to generate test cases that include effective assertion statements poses a challenge. A promising approach is mining or deducing the design requirements of the code under test from historical test data to establish an accurate test oracle.

\begin{tcolorbox}[breakable,width=\linewidth-2pt,boxrule=0pt,top=3pt, bottom=3pt, left=4pt,right=4pt, colback=gray!15,colframe=gray!15]
\textbf{Summary:} For unit test generation, it faces two primary challenges: syntactical correctness and comprehensive semantic coverage. Syntactically, the test code must be compilable and executable. Semantically, it should cover all execution paths and accurately verify the code's intended functionality. 
\end{tcolorbox}

\subsection{Challenges in Code Review}
Code review refers to the process where developers, apart from the coders, meticulously examine the modifications made in a code commit request. 
In traditional workflows, these critical code review activities require significant human effort and call for review outcomes of exceptional quality~\cite{lu2023llama}.
The rapid development of LLMs has brought new opportunities for automating code review with some human intervention and guidance.
However, due to the lack of knowledge base {of code review, which} 
heavily relies on specific requirements and objectives,
complete automation and intelligence in code review are not yet achievable \cite{tufano2024code}. The challenges of LLMs in code review applications primarily manifest in four aspects: (1) {the lack of evaluation and refinement of} high-quality code review comments, (2) construction of code review applications tailored to different types of issues, (3) gap between code review practices in industry and open-source communities, and (4) end-to-end automation of the code review process.

\vspace{0.1cm}\noindent \textbf{(1) {Constructing high-quality code review data by evaluation and optimization.}} Compared to tasks like code generation, existing LLMs lack optimization 
for code review tasks. 
It is necessary to use 
high-quality
code review{-specific} data to 
fine-tune LLMs, enabling them to better adapt to code review tasks. 
However, the quality of code review feedback data varies {in real-world projects.}
For example, the feedback heavily relies on other comments (e.g., ``same as above'' and ``refer to the previous comment''), 
{leading to unclear objectives and expressions \cite{2015-Kononenko}.}
Therefore, it is crucial to {design an evaluation method to select high-quality code review data and propose an optimization approach to refine low-quality code review comments.}

\vspace{0.1cm}\noindent \textbf{(2) Building code review applications for different types of issues.}
In real-world scenarios, code changes submitted for review often involve various issues, including security vulnerabilities, 
coding standard violations, and general lower-priority code formatting issues \cite{survey-cr}. Existing code review applications 
based on LLMs 
{handle all problem types as a generic code review, which} 
causes LLMs to primarily focus on finding {more common problems, such as} standard and formatting issues,
making it difficult to detect more severe problems. 
Furthermore, due to the nature of supervised fine-tuning and the {corresponding relation between a set of code changes and a single code review feedback \cite{DBLP:conf/sigsoft/LiLGDJJMGSFS22},} 
building code review applications for different types of problems is a significant challenge in the code review process.

\vspace{0.1cm}\noindent \textbf{(3) {Mitigating the} gap between code review practices in industry and open-source communities.} 
{Existing code review research related to LLMs} primarily focuses on {open-source} software projects.
However, there are differences between code review practices in open-source communities and industry.
Specifically, developers in the industry are most concerned about security issues that directly impact the developers and the product's viability \cite{2016-Baum},
and open-source community projects involve fewer security issues and more diverse formatting and documentation-related problems \cite{2017-O-Thompson017}. 
Therefore, addressing key industry requirements, acquiring LLMs {post-training based on internal data}, and effectively utilizing prompt engineering {for}
inference are crucial challenges for industry-based code review applications.

\vspace{0.1cm}\noindent \textbf{(4) End-to-end automation of the code review process.} 
Current code review automation technologies mainly focus on automating specific activities within the code review process. 
These existing technologies are primarily based on deep learning and pre-training models{ \cite{DBLP:conf/icse/Tufano23,DBLP:conf/sigsoft/LiLGDJJMGSFS22}, requiring fine-tuning on specific datasets for different downstream tasks and then finishing the whole code review process by stages.}
In the era of LLMs, achieving full automation of the code review process has become an exploratory challenge. 
For example, when a developer proposes a requirement, 
the LLMs can generate problem tickets from the requirement, locate the corresponding code, automatically generate fix solutions, and ultimately achieve end-to-end automation and intelligence from requirement proposal to problem resolution. 

\begin{tcolorbox}[breakable,width=\linewidth-2pt,boxrule=0pt,top=3pt, bottom=3pt, left=4pt,right=4pt, colback=gray!15,colframe=gray!15]
\textbf{Summary}: The challenges in code review applications can be delineated across four aspects: (1) {There is a lack of methods to evaluating and refining real-world code review comments for constructing high-quality code review datasety};
(2) the development of specialized code review tools is constructed to address different types of issues; (3) the gap {in the purpose and process exists} between industrial practices and open-source communities; (4) it lacks an end-to-end automation pipeline within the code review process.
\end{tcolorbox}

\subsection{Challenges in Software Maintenance}

{In modern software maintenance, adopting
LLMs can enhance
modular management and data-driven decision-making capabilities of existing approaches, thereby improving the maintenance
efficiency. However, despite their potential to handle complex systems, they still face many challenges in practical operational scenarios: (1) the complexity of service dependencies in microservice architectures, (2) the lack of high-quality private operational data, and (3) the need for interpretability and high reliability in operational information.}

{\textbf{(1) Complexity of service dependencies in microservice architectures.} In microservice architectures, systems are divided into multiple independently operating service units. This design strategy greatly enhances system modularity and scalability. However, it also leads to fragmented contextual information and increases the complexity of dependencies between service units~\cite{microservice1}. The dispersion of information poses challenges to LLMs because they rely on continuous and consistent contextual information for accurate decision-making and prediction. The dependencies between microservices are complex and dynamic, further limiting a single model's ability to capture the full scope of cross-service operations or transactions. Therefore, to achieve effective operational management in microservice architectures, there is an urgent need to develop innovative technologies to enhance LLMs' ability to integrate dispersed contextual information and improve their understanding of complex inter-service dependencies.}

{\textbf{(2) Lack of high-quality private operational data.} In managing large-scale systems, performance optimization is crucial, especially when handling massive data and high user volumes. However, the insufficient quality and quantity of operational data make LLMs difficult to effectively optimize performance and accurately predict potential system bottlenecks. This issue is particularly prominent in industrial environments which involve highly proprietary data and specific system requirements~\cite{industrial}. These private data, due to sensitivity and confidentiality, are usually not publicly available, further limiting training data sources for LLMs. This confidentiality also leads to uneven data quality, deepening the difficulty for LLMs to perform real-time monitoring and optimization in industrial scenarios.}

{\textbf{(3) Insufficient interpretability and reliability of the results.} 
For ensuring the software reliability~\cite{faustino2022devops}, LLMs-based approaches face unique challenges.
First, LLMs must avoid generating inaccurate or false information when handling complex operations tasks, as this can lead to serious consequences. Also, since operational decisions directly impact system performance, all results must be easy for operation teams to
to understand and verify~\cite{Reliability1}. The opaque nature of LLMs may limit their use in high-reliability environments.}

\begin{tcolorbox}[breakable,width=\linewidth-2pt,boxrule=0pt,top=3pt, bottom=3pt, left=4pt,right=4pt, colback=gray!15,colframe=gray!15]
{\textbf{Summary:} The area of LLMs for software maintenance still faces several challenges, including the complexity of service dependencies in microservice architectures, lack of high-quality private operational data, and need for interpretability and reliability of the operational information.}
\end{tcolorbox}

\subsection{Challenges in Vulnerability Management}

The rapid development of LLMs is bringing new possibilities for building and enhancing vulnerability management capabilities. The major scenarios {of vulnerability management applications} include perception, validation, explanation, analysis, and remediation {for vulnerability scenarios~\cite{DBLP:journals/csur/LeCB23}}. 
{However, LLM-based vulnerability management applications are still in the preliminary validation and exploration stage~\cite{DBLP:conf/apsec/FuTNL23}.} There are several important challenges to overcome before achieving large-scale, efficient, and reliable applications, including (1) the lack of understanding of vulnerability information within the LLMs; (2) the scarcity of high-quality vulnerability analysis data; 
(3) the complexity of vulnerability context information.
, which {need to extract adequate slices as input and consider the window size of LLMs.}

\vspace{0.1cm}\noindent \textbf{(1) Lack of understanding of vulnerability information.}
LLMs are typically trained on open-source data from various internet sources~\cite{DBLP:journals/corr/abs-2402-15100}. However, the availability and quantity of open-source data containing vulnerability information are limited~\cite{DBLP:conf/crisis/GhaniLKAS13}, resulting in an insufficient understanding. 
{It} impacts the accuracy and reliability of vulnerability management applications.
{Therefore, native LLMs require extensive vulnerability data to effectively develop specialized vulnerability management, highlighting the importance of data-driven training in the construction of corresponding applications.}

\vspace{0.1cm}\noindent \textbf{(2) Scarcity of high-quality vulnerability analysis and explanation data.}
The strength of generative models lies in providing detailed analysis and explanations for problems. However, 
building vulnerability management applications based on LLMs 
{lacks the} high-quality vulnerability explanation and analysis data. 
{It is due to that} open-source vulnerability information datasets mainly provide information such as code patches, CVE IDs, and CWE types~\cite{DBLP:conf/promise/BhandariNM21}, but 
{it is less accessible for developers to understand and effectively address vulnerabilities.}
Therefore, obtaining high-quality vulnerability analysis and explanation data 
{for } vulnerability management applications remains an important challenge.


\vspace{0.1cm}\noindent \textbf{(3) Complexity of vulnerability context information.}
It is also essential to provide sufficient vulnerability context information in the prompt~\cite{DBLP:journals/corr/abs-2404-15596}. This context includes the code location, vulnerability type, 
and detailed patch information. However, these pieces of information can be lengthy, potentially exceeding the window size of existing LLMs and leading to the loss of important information. LLMs may struggle to consider all aspects of vulnerabilities simultaneously, resulting in inaccurate assessments of severity and impact. 

\begin{tcolorbox}[breakable,width=\linewidth-2pt,boxrule=0pt,top=3pt, bottom=3pt, left=4pt,right=4pt, colback=gray!15,colframe=gray!15]
\textbf{Summary}: Vulnerability management applications have the following challenges:
{First, to better understand vulnerability information, LLMs need to be enhanced in their interpretability by training on extensive vulnerability data. Second, high-quality vulnerability analysis data are scarce. It is important to obtain high-quality
vulnerability analysis and explanation data for vulnerability management applications.}
Finally, current LLMs may not fully capture the extensive and intricate context surrounding vulnerabilities, which {need to extract adequate slices as input while
considering the limited window size.}
\end{tcolorbox}

\subsection{Challenges in Data, Training, and Evaluation}
In this section, we discuss the challenges related to data construction, model training, and performance evaluation, which could impact the integration of LLMs into the software development process in practice.

\noindent\textbf{(1) High-Quality Code Data.} 
The LLMs have successfully been applied in code generation, which attracts considerable research attention toward developing LLMs specialized in software engineering for downstream tasks. Despite the rapid evolution and iteration of LLMs, their core architecture is primarily based on the Transformer model. In this scenario, high-value data sources and high-quality pre-training data have become key factors for enhancing LLM performance {\cite{kocetkov2022stack}} and could become companies' core commercial assets. Therefore, constructing high-quality code data presents numerous challenges.

During the pre-training phase of code LLMs, the scale and diversity of data are two crucial factors affecting model performance. Open-source code communities (such as GitHub and Stack Overflow) have become the primary data sources. However, the quality of data from these communities varies greatly. Publicly available datasets in software engineering often lack rigorous quality assurance. Additionally, the current practices for source code cleaning typically rely on heuristic text cleaning rules {\cite{zhang2024data}}, which mainly address format issues but fall short of ensuring data consistency, diversity, accuracy, and completeness. These practices also do not effectively utilize the structural and semantic features of code. \textit{Therefore, the challenge of creating a large-scale, diverse, and high-quality code dataset during the pre-training phase and defining standards for high-quality code data remains unresolved.} 

During the fine-tuning phase for various downstream tasks, high-quality and labeled data is essential for enhancing the performance of code LLMs. Key challenges include identifying relevant code snippets and their contexts within extensive code datasets for effective fine-tuning, as well as accurately labeling data while minimizing manual labor costs. Furthermore, it is critical to refine existing datasets for downstream tasks to ensure their quality and expand these datasets to encompass a broader range of software engineering tasks. Although advanced models, such as the GPT family, offer support for data annotation {\cite{ding2022gpt}}, their effectiveness in generating consistent and robust code remains challenging due to several factors. Addressing these issues is critical for improving the utility and accuracy of code LLMs in solving different software engineering tasks.

\noindent\textbf{(2) Exhaustive Training.} Training LLMs present challenges such as high costs and poor stability. The lack of advanced engineering capabilities for training further exacerbates these issues. Companies like Meta and OpenAI have expressed concerns about the excessive proportion of computational resources dedicated to training debugging, which further drives up the cost of training large models. Existing model training dashboards, such as TensorBoard~\cite{tensorboard}, only provide basic training features and offer limited assistance to model training personnel in debugging models. Therefore, the training process requires advancements in deployment, configuration, bug localization, and process monitoring.

\noindent \textbf{(3) Comprehensive evaluation.} Evaluating code LLMs involves two critical phases: the \textit{model-selection phase} and the \textit{post-selection phase}, where they focus on optimizing research and development efficiency in practical applications. Each phase introduces different challenges.

During the model-selection phase, evaluation typically relies on benchmark datasets. However, existing studies indicate that the performance of LLMs on established benchmarks (e.g., HumanEval and MBPP) has reached near-limit bottlenecks as these models evolve, making it difficult to conduct comprehensive evaluations. Recently, evaluation datasets derived from real-world development projects have been released {\cite{feng2024complexcodeeval}}, yet their creation demands considerable manual annotation and verification effort. Challenges in this phase include setting evaluation standards, efficiently constructing high-quality benchmark datasets, and ensuring these datasets comprehensively represent the scope of evaluation.

The evaluation of code LLMs in real-world usage presents distinct challenges. Benchmark-based scores (such as hit rate, coverage rate, and edit distance) do not accurately capture the models' practical utility to developers. Developing evaluation frameworks that mimic real development contexts and feeding the evaluation of generated content into the model training phase to identify training or data issues represent advanced requirements and remarkable challenges in model evaluation.

\begin{tcolorbox}[breakable,width=\linewidth-2pt,boxrule=0pt,top=3pt, bottom=3pt, left=4pt,right=4pt, colback=gray!15,colframe=gray!15]
\textbf{Summary:} For the training data, it needs to create a large-scale, diverse, and high-quality code dataset to boost code LLMs' effectiveness. Training LLMs also presents high costs and poor stability due to the lack of advanced engineering capabilities for training debugging. Regarding the evaluation part, it is essential to establish frameworks that replicate real-world development settings, allowing for a thorough and precise measurement of the models' practicality.
\end{tcolorbox}

\section{Related Work}\label{sec:literature}

\subsection{LLM-based Software Design}

In recent works, LLMs have been used in various subtasks of requirement engineering, including anaphoric ambiguity treatment, requirement classification, requirement terminology recognition, coreference detection, and traceability automation.

For anaphoric ambiguity treatment, Ezzni's empirical study~\cite{ezzini2022automated} verifies that SpanBERT excels at coreference explanation but has limited performance in ambiguity detection. 
Moharil and Sharma~\cite{moharil2023tabasco} introduce a toolkit called TABASCO to detect and recognize intra- or inter-domain ambiguity. It utilizes BERT for representation computation and clustering algorithms for analysis, proving the effectiveness of BERT in ambiguity detection. 
Sridhara et al.~\cite{sridhara2023chatgpt} applies ChatGPT to multiple SE tasks. In requirement engineering, ChatGPT successfully identifies all antecedents in the requirements for coreference detection, preliminarily verifying the applicability of ChatGPT to enhance the clarity of requirement expressions and eliminate ambiguity to some extent.

During project initialization, requirements often need to be categorized, such as distinguishing security-related requirements from others. They may also be divided based on relevance to functionality. Moreover, specific terminologies and entities in requirements may have different meanings in different contexts. 
NoRBERT~\cite{hey2020norbert} is a fine-tuned BERT that takes advantage of BERT's strength in transfer learning and achieves huge improvements in identifying functional and non-functional requirements compared to previous approaches. 
PRCBERT~\cite{luo2022prcbert} benefits from prompt tuning, which enhances BERT's classification ability, outperforms NoRBERT and demonstrates notable capabilities in zero-shot inference on other datasets. 
Moharil et al.~\cite{moharil2022identification} extract terminologies and synonyms from requirements with BERT and clustering algorithms, showcasing sample sentences from the corpus to clarify the meanings of terminologies within specific contexts. 
DeepCoref~\cite{wang2020deep} employs fine-tuned BERT and a Word2Vec model to represent entities and predict associations between two entities.

Furthermore, LLMs are widely adopted to solve other requirement engineering tasks, like traceability automation, which establishes and maintains relationships among requirements, code, test samples, and other elements in the software system. 
T-BERT~\cite{lin2021traceability} is a framework that can create tracing links between code and natural language descriptions. Due to limited training data, traditional deep learning models often underperform pre-trained models in this task. 
Poudel et al.~\cite{poudel2023leveraging} release Sat-BERT and DSat-BERT for requirement satisfaction assessment, providing more accurate results than information retrieval approaches. 
Ronanki et al.~\cite{ronanki2022chatgpt} use ChatGPT to evaluate the quality of user stories, which shows that the judgments from ChatGPT align well with human evaluations. 
SpecGen~\cite{ma2024specgen} leverages LLMs to verify the formal specifications of programs. 
It first generates specifications for a given program and employs multiple mutations to refine the generated specifications. 
Xie et al.~\cite{xie2023impact} conduct an empirical study on utilizing LLMs to generate software specifications with few-shot learning, which evaluates the performance and costs of 15 LLM-generated specifications and emphasizes the importance of prompt design and domain knowledge. 
Endres et al.~\cite{endres2023formalizing} convert informal natural language descriptions into assertions with LLMs, and verify the correctness of generated assertions, which can help identify defective code. 
Zhang et al.~\cite{zhang2024experimenting} propose a new development practice, AISD, which first accepts user requests and then generates detailed applicable scenarios and prototype system designs before proceeding to generate the system implementation.

\subsection{LLM-based Code Assistance and Generation}
Code generation can be applied to various scenarios, including code completion, code search, bug fixing, code review, and automatic testing. These applications aim to enhance development efficiency, reduce human errors, and assist developers in implementing new features rapidly. In recent years, code LLMs have been widely applied to code generation tasks. These models adopt different architectures, including CodeBERT \cite{25}, PLBART \cite{26}, and CodeGPT \cite{27}. They are pre-trained on code corpora to gain a deep understanding of code syntax, semantics and common structures. To boost their understanding of code complexity, some innovative methods integrate structural representation. GraphCodeBERT \cite{28} incorporates graph-based representation on top of CodeBERT, while CodeT5 \cite{29} combines the Encoder-Decoder paradigm with code structure. These enhancements aim to provide more fine-grained understandings of code relations and dependencies for the model in addition to syntax information.

Open-source communities have provided abundant resources for code generation research, such as open-source repositories, datasets, and tools. For instance, GitHub, Stack Overflow and other platforms offer plenty of code samples, which are used to train and evaluate LLMs. CodeX \cite{1} and CodeGen \cite{13} are constructed with billions of parameters, demonstrating state-of-the-art performance on code generation tasks, capable of assisting the developers to generate code and improve programming efficiency. The success of CodeX prompts the development of similar models like StarCoder \cite{22} and CodeLlama \cite{codellama}. Developers can also use ChatGPT and GPT-4 to generate code, which has been proven effective in such fields.

However, instead of developing new LLMs, existing studies usually exploit LLMs as black-box tools. 
They integrate many techniques of In-Context Learning (ICL), multi-agent systems, integrating traditional methods, user interaction, and Chain-of-Thought (CoT) reasoning to improve the effectiveness of LLMs in solving different tasks.

Jiang et al.~\cite{jiang2023self} propose a two-stage code generation pipeline based on ICL, i.e., \textit{planning} and \textit{implementation}. 
This pipeline allows LLMs to plan the solution steps according to users' intent before generating code.
Li et al.~\cite{li2023towards} propose a new framework, AceCoder, which retrieves similar code as cases so that LLMs can learn algorithms and API knowledge. 
It then instructs LLMs to generate test cases and APIs to ensure that they understand the requirements before generating the code. 
LAIL~\cite{li2023large} trains an auxiliary retriever to fetch examples for ICL, which first estimates the probability of generating the sample programs given the requirements and inputs with LLMs, then labels samples as positive or negative. It initiates contrastive learning for the retriever to learn LLM preferences, so it can select the optimal examples for ICL. 
On the other hand, as inserting print statements is effective for debugging, Hu et al.~\cite{hu2024leveraging} utilize this debugging approach and instruct LLMs to insert print statements with ICL, execute the code locally, and ask LLMs to refine the code based on the outputs.

Multi-agent systems play different roles in accomplishing tasks collaboratively. 
These roles often mimic real-world development processes. 
Dong et al.~\cite{dong2023self} point out that code generation research should not be confined to programming alone, and suggest adopting multiple LLM agents to simulate different development team roles, such as analysts who break down user requirements, developers who generate or update code, and testers who test the code from various aspects.
AgentCoder~\cite{huang2023agentcoder} adopts three agents, including a developer, a test designer, and a tester. The developer agent generates and refines the code based on tester feedback, the test designer agent generates test cases given the code, and the tester agent executes the test and offers feedback. 
LCG~\cite{lin2024llm} simulates real-world development processes with multiple agents, including requirement engineers, architects, developers, and testers, and improves the development performance of GPT-3.5 with CoT and prompt engineering techniques. 
Similarly, MapCoder~\cite{islam2024mapcoder} employs four agents, each responsible for collecting relevant cases, planning, generating code, and debugging, respectively.

As LLM struggles with repository-level code generation, some researchers combine traditional tools to bridge the gap. 
CodePlan~\cite{bairi2024codeplan} studies repository-level code generation, constructing planning graphs based on dependency analysis and adaptive planning, with edit positions as nodes and edit orders as edges. 
Information from these graphs is integrated into the prompts used to generate code. 
The code is then merged into the repository, causing an update to the planning graphs as the process continues iteratively.
ToolGen~\cite{wang2024teaching} integrates auto-completion tools to resolve the dependency issues in repository-level code generation. 
It first enhances the code in the repository with special tokens for LLM fine-tuning alongside documentation. During token sequence generation, the auto-completion tool produces the correct completion entry when the LLM generates a particular token. 
Zhang et al.~\cite{zhang2023toolcoder} release ToolCoder, which fine-tunes LLMs using a ChatGPT-augmented dataset with tool-use information. 
During inference, an API search tool is integrated to offer suggestions on API selection. 
Zhang et al.~\cite{zhang2024codeagent} propose CodeAgent, integrating five tools for tasks like information retrieval and code symbol navigation, as well as four agent strategies like OpenAI's function calling, to help LLMs generate code. 
However, not all studies report state-of-the-art performance for LLMs. For example, Bochenek~\cite{bochenek2023can} tries to use ChatGPT to complete Java code templates. 
In conclusion, the author claims that the results are neither predictable nor reproducible, and traditional template-filling approaches perform better.

Some studies investigate the interactions between users and LLMs. 
Fakhoury et al.~\cite{fakhoury2024llm} propose a novel framework, TiCoder, to ensure the correctness of generated code. 
It first generates code and test cases with LLMs, tests the code locally, asks LLMs to select the best test cases based on the test results, and finally requests the users to evaluate the correctness of the code, test cases, and test results. 
Unsatisfactory test cases are then pruned, and the framework updates the code and test cases iteratively. 
Considering the lack of domain knowledge, Yan et al.~\cite{yan2024intelliexplain} propose IntelliExplain, which requires LLMs to explain the users' intent in detail in natural language and avoid code generation until the user agrees with the LLMs' explanation.

To mitigate misalignment between code and specifications in previous CoT methods, Tian et al. ~\cite{tian2023test} design a test case-driven CoT approach named TCoT, which lets LLMs understand specifications through test cases before generating the code. 
Sun et al.~\cite{sun2024enhancing} propose CodePLAN, which uses multi-task learning to distill the ability of code generation and planning from larger LLMs to smaller ones.

Other studies examine code generation from perspectives beyond the correctness of the code. 
Nguyen et al.~\cite{nguyen2024gptsniffer} construct GPTSniffer with CodeBERT, which helps to identify whether the code is generated by AI tools like ChatGPT. 
Huang et al.~\cite{huang2023bias} investigate whether LLM-generated code contains biases related to age, race, and gender. 
It uses GPT-4 to detect biases and utilizes few-shot learning to reduce them. 
Niu evaluates the efficiency of generated code on existing datasets and explores how prompt strategies affect code efficiency.

\subsection{LLM-based Software Testing}
In software testing, LLMs are typically used to generate test cases for automatic testing, construct assertions, produce test inputs, analyze defects, and help debug and fix the code. For example, LLMs can generate expected test cases by understanding code contexts and structures \cite{30}, to improve test coverage and effectiveness. Furthermore, LLMs can generate accurate assertions \cite{31}\cite{46} as part of test cases, to validate whether the behavior of the software meets the expectations. On the level of system testing, LLMs can generate various test inputs, which is essential for testing user interfaces and functionalities of mobile applications.


Recently, various approaches aim at generating unit tests with LLMs, which is often associated with pre-training or fine-tuning. 
Alagarsamy et al.~\cite{alagarsamy2024a3test} propose A3Test, a framework inspired by domain adaptation, which transfers assertion generation knowledge into test code generation. 
They pre-train LLMs with methods to be tested and assertion statements, equipping LLMs with more knowledge about assertions. 
The pre-trained models are then fine-tuned for test code generation, allowing them to better understand the relationship between the methods to be tested and their corresponding test code. 
Similarly, Hashtroudi et al.~\cite{hashtroudi2023automated} pre-train LLMs with existing developer-written test code, helping models adapt to the new project domains and generate human-readable unit tests. 
Rao et al.~\cite{rao2023cat} introduce CAT-LM, which uses a novel pre-train signal to map code to test files and train a 270M GPT-style language model. 
Steenhoek et al.~\cite{steenhoek2023reinforcement} release RLSQM, which scores programs based on static analysis and embeds the quality information into LLMs, including factors such as the presence of assertion statements, whether the methods to be tested are invoked, and the existence of method signatures.

With the development of LLMs, they can exhibit satisfying performance without further pre-training or fine-tuning. Therefore, many recent researches instead focus on prompt strategies to improve LLMs' ability to handle context and details. 
Xie et al.~\cite{xie2023chatunitest} propose ChatUniTest, adopting a generate-verify-repair framework, which analyzes project information and extracts key data to create a context embedded with methods to be tested as well as their dependencies. 
After the LLMs receive the prompt and generate feedback, ChatUniTest extracts test code from the outputs, verifies the code, and fixes compilation errors with either rule-based or LLM-based approaches. 
Other works generate unit tests with additional documentation. Plein et al.~\cite{plein2024automatic} suggest generating test code that can reflect real-world, complex usage scenarios, which exposes faulty program behavior. As such scenarios are often described informally in defect reports, they consider such reports as valid inputs to generate fault-triggering test code, unlike previous works focusing on generating random test inputs.

Researchers also develop many novel methods to generate test inputs or test oracles with LLMs. These inputs validate mobile apps, web applications, and deep learning frameworks.
To automate the process of mobile app testing, researchers propose different test input generation approaches with LLMs. 
Yoon et al.~\cite{yoon2023autonomous} propose DROIDAGENT, which sets goals with LLMs given the Android app and achieves the goals by interacting with the apps. 
The evaluation of DROIDAGENT shows that it can engage in deeper interaction with apps to cover more features. 
Liu et al.~\cite{liu2023fill} propose QTypist, which uses LLMs to generate input text for mobile app GUIs, solving the challenge of generating varied and semantically correct GUI inputs. 
The evaluation shows that QTypist successfully improves the test pass rate and covers more app activities and pages. 
Liu et al.~\cite{liu2024testing} proposed InputBlaster to generate abnormal text inputs for mobile apps with LLMs. 
It creates multiple test generators, each capable of generating abnormal inputs based on LLM-generated mutation rules.

In web test automation, Kim et al.~\cite{kim2024leveraging} propose RESTGPT, generating test inputs for REST API with LLMs, which overcomes the limitations of previous methods. It struggles to extract rule types from unstructured natural language and improves accuracy. 
Another issue with LLM-generated programs is the failure to cover corner cases. 
To solve this issue, Deng et al.~\cite{deng2023large} release FuzzGPT, which preprocesses LLMs and synthesizes fault-triggering test inputs for fuzzing. FuzzGPT automatically extracts historical knowledge of fault-triggering programs with LLM capabilities. 
Evaluation results prove FuzzGPT's effectiveness in fuzzing and detecting more errors.

In other domains, researchers also propose innovative methods for LLM-based test input generation. 
For example, Baudry et al.~\cite{baudry2024generative} highlight the ability of LLMs to generate pseudo-test data and successfully generate data aligning well with cultural contexts and suitable for testing, as well as code compatible with test input generation tools. 
Xia et al.~\cite{xia2024fuzz4all} propose Fuzz4All, a generic fuzzing tool using LLMs as an engine to generate and mutate test inputs. 
Fuzz4All supports multiple programming languages and scenarios, such as C/C++ compilers, solvers, tools in Go, Java compilers, and quantum computation platforms.

Test oracles can validate the correctness of test results. 
In traditional approaches, test oracles are usually derived from specifications. 
However, manual generation is labor-intensive, while automatic generation proves to be extremely challenging. In recent years, researchers have started to use LLMs to help generate test oracles. Metamorphic testing (MT) has emerged as a successful solution to automated testing and test oracle generation. MT relies on metamorphic relations (MRs), which describe the relationship between test inputs and outputs. 
Shin et al.~\cite{shin2024towards} utilize LLMs to infer executable MRs (EMRs) from requirements, guiding them to understand software specifications and domain-specific languages (DSL) for EMRs to generate the test oracle. 
Zhang et al.~\cite{zhang2023automated} explore applying ChatGPT to generate MRs automatically and evaluate their quality for auto-driving systems (ADSs), demonstrating the effectiveness of the proposed approach, which significantly reduces manual effort in MR generation. 
Tsigkanos et al.~\cite{tsigkanos2023large} design an LLM-based process to extract variables from user manuals of scientific software, achieving an accuracy of 0.87 and successfully extracting 61.8\% partially-matching variables as well as 34.7\% exact-matching variables. 
Hyun et al.~\cite{hyun2024metal} introduce METAL, promoting systematic evaluation of LLM quality with MT. 
METAL can automatically generate hundreds of MRs from templates covering various qualitative properties and tasks. Furthermore, they introduce new metrics, which combine attack success rate and semantic quality, to better evaluate the effect of MT.



\subsection{LLM-based Code Review}
The applications of LLMs in code analysis and review are a trending topic in the current field of artificial intelligence. These models utilize deep learning techniques to learn a large amount of code data so that they can understand and generate programs with complicated structures. These LLMs play an important role in software development.

In code analysis, researchers have developed multiple models to understand code diffs and reviews. Universal models like GPT-3 display outstanding performance in several natural language processing tasks, but on the other hand, might be less accurate in code analysis compared to specialized models. Specialized models, such as CodeBERT and GraphCodeBERT, are specifically built on understanding code semantics, pre-trained on large code corpora, and can better capture code syntax and semantics.

In code review, LLMs are mainly applied to review code and detect flaws automatically. For instance, DeepCode \cite{36} is a code review tool powered by machine learning, which can detect security vulnerabilities and performance issues to help developers enhance the quality of code. CodeReviewer~\cite{DBLP:conf/sigsoft/LiLGDJJMGSFS22} is a pre-trained model tailored specifically for the code review scenario, which is trained on a large-scale dataset of real-world code changes and code reviews from open-source projects, helping to better understand code diffs and reviews.
Tufano et al. ~\cite{tufano2021towards,tufano2022using} approach code review from the perspectives of both the developer and the reviewer, and train T5 for two subtasks of code review. The first subtask requires the model to refine the given code, while the second requires the model to edit the code based on a review comment. 
AUGER~\cite{li2022auger} is a framework utilizing the pre-trained model CodeTrans, a variant of T5. It labels lines of code relevant to the review comment, performs cross-pre-training between functions and comments, and fine-tunes for comment generation. 
AutoTransform~\cite{thongtanunam2022autotransform} addresses new tokens with Byte Pair Encoding (BPE) and adopts Transformer in a machine translation manner to edit the code after review.
Zhou et al.~\cite{zhou2023generation} study previous generative code review methods and conclude that: 1. They are based on different datasets, making comparisons difficult; 2. There are few studies targeting CodeT5; 3. Exact match (EM) is often the primary metric, while other metrics are overlooked. To address these issues, they adopt CodeT5, introduce a new metric called edit process (EP), and compare their approach with previous works.

Llama-Reviewer~\cite{lu2023llama} uses Llama in code review, utilizing parameter-efficient fine-tuning (PEFT) to reduce resource consumption while maintaining high performance. As a result, even the smallest Llama-6.7B performs comparably to previous methods.
Wen et al.~\cite{wen2024automatically} use GPT-3.5 for code analysis and review. They first extract relevant code with static tools, bug reports, and program dependencies, before constructing formal prompts that incorporate domain knowledge and representative examples, and prompting the LLM to evaluate the accuracy of warnings from static tools, greatly reducing false positives. 
Pornprasit et al.~\cite{pornprasit2024gpt} also use GPT-3.5 for code review, investigating the performance impact of few-shot learning, role specification, and model fine-tuning, and compare the results with previous approaches. 
Tufano et al.~\cite{tufano2024code} study whether current methods are sufficient for comment generation and comment-based code editing, examining the types of samples where these methods succeed or fail and comparing them with ChatGPT.

\subsection{LLM-based Software Maintenance}
{As software systems grow larger, the maintenance becomes more complex and critical. The introduction of LLMs brings new opportunities to software maintenance, such as failure identification~\cite{maintain1,maintain2,maintain3,maintain4}, root cause analysis~\cite{maintain5,maintain6,maintain7,maintain8,maintain9}, and postmortem analysis~\cite{maintain10,maintain11,maintain12}.}

{For failure identification, researchers use LLMs for log parsing and anomaly detection, addressing challenges in semantic understanding and efficiency. For example, Huo et al.~\cite{maintain1} first consider semantic information in log parsing. Le et al.~\cite{maintain2} try to use ChatGPT for log parsing and initially confirm the ability of LLMs in this area. However, LLMs have issues such as long response times and high computational costs when processing a large number of logs. To address this, Jiang et al.~\cite{maintain3} propose an adaptive caching mechanism that caches log templates to avoid repeated requests and improve parsing efficiency. Ma et al.~\cite{maintain4} combine external knowledge to enhance log analysis.
For root cause analysis, LLMs help understand system events and diagnose faults by combining code and log data for automatic analysis and recommendations. Ahmed et al.~\cite{maintain5} use LLMs to automatically output fault root causes and suggest mitigation measures. Wang et al.~\cite{maintain6} propose a tool-learning method. Zhang et al.~\cite{maintain7} introduce a confidence evaluation framework to alleviate issues like model hallucination. Jin et al.~\cite{maintain8} use LLMs for fault impact assessment and summary generation. Chen et al.~\cite{maintain9} develop RCACopilot to achieve an end-to-end fault handling process.
For postmortem Analysis, LLMs assist in summarizing incident tickets, improving data labeling efficiency, and promoting fault prediction. Dogga et al.~\cite{maintain10} propose fault classification rules and attempt automatic classification. Ganatra et al.~\cite{maintain11} apply LLMs to the fault analysis pipeline. The Huawei Cloud team~\cite{maintain12} propose a hierarchical representation method to achieve automatic classification and prediction.}

{Overall, the application of LLMs in software maintenance improves efficiency, simplifies complex tasks, and reduces the time for repairing. However, it also brings challenges like complexity of service dependencies and model interpretability that need to be addressed.}

\subsection{LLM-based Vulnerability Management}
Recently, many studies have proposed to manage vulnerabilities, including vulnerability detection~\cite{DBLP:journals/tse/ZhangLHXL23/EPVD, wen2024scale, wen2024when, wen2023vulnerability}, vulnerability assessment~\cite{pan2024towards}, and vulnerability awareness~\cite{10.1145/3540250.3549156, zhou2023colefunda}.

Vulnerability detection aims to detect vulnerabilities in the source code~\cite{wen2024when, DBLP:journals/tse/WenGYLTJW24}. Automated code vulnerability detection by using LLMs has gained increasing attention in recent years~\cite{DBLP:journals/jss/LuJCPC24, DBLP:conf/tpsisa/HuHIT023}. 
EPVD~\cite{DBLP:journals/tse/ZhangLHXL23/EPVD} proposes an execution path selection algorithm and adopts a pre-trained model to learn the path representations. SVulD~\cite{DBLP:journals/corr/abs-2308-11237/svuld} constructs contrastive paired instances and uses the pre-trained model to learn distinguishing semantic representations.

To improve early sensing of vulnerabilities, many studies are proposed to make aware of vulnerabilities from issue reports~\cite{10.1145/3540250.3549156} or silent fixes~\cite{zhou2023colefunda}.
Open source software usually adopts a vulnerability disclosure policy which causes the majority of vulnerabilities to be fixed silently.
Identifying silent fixes and their corresponding explanations is essential to sense vulnerabilities. 
Zhou et al.~\cite{zhou2023colefunda} propose an approach named ColeFunda, consisting of a Contrastive Learner and FunDa, a novel approach for Function change Data augmentation. ColeFunda integrates contrastive learning and BERT models to improve its effectiveness in identifying different vulnerabilities.
To sense vulnerabilities early, Pan et al.~\cite{10.1145/3540250.3549156} augment the BERT model with a memory component, which stores the external vulnerability knowledge from CWEs.



\section{Threats to Validity}\label{sec:threats}
One threat is the completeness of the challenges. The summarized challenges are based on the discussion among 24 participants, which might not cover all the current challenges of LLM4SE. In this paper, to ensure the integrity of the challenges, the discussion was designed to encompass all the processes in SDLC and included six thematic sessions. 
Each session lasted for around four hours to ensure that the discussion was as thorough as possible.

Another threat is related to the representativeness of the participants. Our solution is increasing diversity, e.g., 17 academic researchers and 7 industry practitioners. 
We believe we have made this threat have minimal impact on the results of our investigation.


\section{Conclusions}\label{sec:con}
LLMs are bringing significant changes to the SE field and are applied to various SE tasks, such as code generation, code review, and vulnerability management.
However, a comprehensive understanding of the challenges of LLMs on SE is still in the early stages.
This paper highlights the challenges and opportunities in LLM4SE.
We present seven challenges summarized in the ``9th \xiuhu''. These challenges include various aspects of the software development life cycle, e.g., requirements, coding, testing, and code review.
This paper also highlights challenges in building LLMs, including data preparation, training, and evaluation. We provide a research roadmap of LLM4SE and outline promising future research directions.



\bibliographystyle{ACM-Reference-Format}
\bibliography{software}


\end{document}